\documentclass[twocolumn,prb,superscriptaddress,preprintnumbers,amssymb]{revtex4-1}

\usepackage{graphicx}% Include figure files
\usepackage{dcolumn}% Align table columns on decimal point
\usepackage{bm}% bold math

\begin{document}

%\preprint{final version}

\title{Structure and stability of small H clusters on graphene}

\author{\v{Z}eljko \v{S}ljivan\v{c}anin \footnote{Electronic mail: zeljko@vinca.rs}}
\affiliation{Interdisciplinary Nanoscience Center (iNANO) and Department 
of Physics and Astronomy, Aarhus University, Ny Munkegade 120, building 1520, 
DK-8000 {\AA}rhus C, Denmark}
\affiliation{Vin\v{c}a Institute of Nuclear Sciences (020),
P.O.Box 522, RS-11001 Belgrade, Serbia}
%\email{zeljko@vinca.rs}
\author{Mie Andersen}
\affiliation{Interdisciplinary Nanoscience Center (iNANO) and Department 
of Physics and Astronomy, Aarhus University, Ny Munkegade 120, building 1520, 
DK-8000 {\AA}rhus C, Denmark}
\author{Liv Hornek{\ae}r}
\affiliation{Interdisciplinary Nanoscience Center (iNANO) and Department 
of Physics and Astronomy, Aarhus University, Ny Munkegade 120, building 1520, 
DK-8000 {\AA}rhus C, Denmark}
\author{Bj{\o}rk Hammer}
\affiliation{Interdisciplinary Nanoscience Center (iNANO) and Department 
of Physics and Astronomy, Aarhus University, Ny Munkegade 120, building 1520, 
DK-8000 {\AA}rhus C, Denmark}

\date{\today}

\begin{abstract}
The structure and stability of small hydrogen clusters adsorbed on
graphene is studied by means of Density Functional Theory (DFT)
calculations. Clusters containing up to six H atoms are investigated
systematically -- the clusters having either all H atoms on one side of
the graphene sheet (\textit{cis}-clusters) or having the H atoms on both
sides in an alternating manner (\textit{trans}-cluster). The most
stable cis-clusters found have H atoms in ortho- and para-positions with respect
to each other (two H's on neighboring or diagonally opposite carbon
positions within one carbon hexagon)
while the most stable trans-clusters found have H atoms in ortho-trans-positions
with respect to each other (two H's on neighboring carbon
positions, but on opposite sides of the graphene). Very stable
trans-clusters with 13-22 H atoms
were identified by optimizing the number of H atoms in ortho-trans-positions 
and thereby the number of closed, H-covered carbon hexagons. 
For the cis-clusters, the associative
H$_2$ desorption was investigated. Generally, the desorption with the lowest
activation energy proceeds via para-cis-dimer states, i.e.\ involving
somewhere in the H clusters two H atoms that are positioned on
opposite sites within one carbon hexagon. H$_2$ desorption from
clusters lacking such H pairs is calculated to occur via hydrogen
diffusion causing the formation of para-cis-dimer states. Studying the
diffusion events showed a strong dependence of the diffusion energy
barriers on the reaction energies and a general odd-even dependence on
the number of H atoms in the cis-clusters.
\end{abstract}


\maketitle
\section{Introduction}
The interaction of H atoms and molecules with carbon based materials like
graphite, single wall carbon nanotubes (SWCNT) and graphene has attracted
considerable interest during the last two decades, since these systems are of interest in fields as diverse as astrochemistry, hydrogen storage and nanoelectronics.\\
The understanding of H chemisorption on graphitic materials and formation of
H$_2$ molecules at low H atom densities is highly relevant for
interstellar chemistry \cite{Astrophysics}. H$_2$ formation in interstellar
dust and molecular clouds is expected to occur via recombination of H atoms adsorbed on
interstellar dust grain surfaces. Since carbonaceous grains are abundant in the
interstellar medium, particular attention is focused on H chemisorption on
graphite surfaces \cite{Cazaux2004}.\\
Most of the recent efforts regarding H adsorption on carbon-based materials
have been stimulated by the possibility of using carbon nanostructures as a
hydrogen storage medium \cite{Schlapbach2001}.
Earlier reports on H storage of $\sim$7.4 wt \% in nanostructured graphite
\cite{Schlapbach2001} and 14 wt \% in SWCNTs \cite{Dillon2001} have not been
confirmed by new experiments that rather indicate that the H storage capacity at room
temperature is 3.8 wt \% in graphite nanofibers \cite{Lueking2004} and
$\sim$5.1 \% in SWCNTs \cite{Nikitin2005}. The precise mechanism of H adsorption
in these structures is unknown since these values are well above the
estimated limit of 1\% for H$_2$ physisorption at room temperature
\cite{Ritschel2002}. For graphene a theoretical maximum storage capacity of 7.7 wt \% is reported \cite{Sofo2007}.\\
Recent fabrication of graphene has generated an enormous number of new studies
focused on the prospects of using graphene as a key material in post silicon
electronics \cite{Novoselov2005,Geim2007,Castro2009}.
The high electron mobility can be exploited for construction of ballistic
transistors operating at frequencies unreachable with current
semiconductor materials \cite{Lin2010}. Yet, for applications in nanoelectronics efficient
methods and tools for tailoring the electronic properties of graphene are required.
One of the most promising routes towards the opening of a band gap in graphene is based
on its hydrogenation \cite{Duplock2004,Sofo2007,Elias2009,Liv2010}. Hence, an atomic scale
description of the binding mechanism and stability of small hydrogen
structures formed on graphene will enhance efforts aimed at full control over
graphene band gap engineering with deposited H atoms.\\
The chemisorption of H monomers on graphite has been studied extensively both
experimentally and theoretically
\cite{Neumann1992,Jeloaica1999,Sha2002,Zecho2002b}. H dimers on
graphite or graphene have also been thoroughly studied in recent years
\cite{Zecho2002b,Hornekaer2006,Hornekaer2006b,Roman2007,Ferro2008,Casolo2008,Boukhvalov2008,Zeljko2009}. According
to experiments, bigger structures will form when graphite or graphene
are exposed to higher H doses \cite{Hornekaer2006b,Hornekaer2007,Elias2009,Balog2009,Liv2010}.
These structures, highly relevant for possible technological applications,
are much less investigated than dimers. Only recently, Casolo
{\it et al.} \cite{Casolo2008},
Roman {\it et al.} \cite{Roman2009}, Ferro {\it et al.} \cite{Ferro2009} and Khazaei {\it et al.}\cite{Khazaei2009}
used Density Functional Theory (DFT) to calculate the energetics of hydrogen 
clusters with three and four atoms, as well as one structure with six atoms on 
a graphene sheet. Cuppen {\it et al.} \cite{Cuppen2008} studied the formation of hydrogen clusters using Kinetic Monte Carlo methods. Luntz and co-workers \cite{Luntz2009} studied desorption of D$_2$ molecules from D clusters adsorbed on graphite, combining experimental and theoretical
methods. A comprehensive list of tetramer structures and the recombination
pathways was produced based on measured desorption density and DFT 
calculations  \cite{Luntz2009}.
In the present paper we apply DFT to systematically study small hydrogen clusters composed of
three to six H atoms on graphene. Their structure and
stability against H diffusion and H$_2$ recombination are determined at
the atomic scale. Two types of clusters are considered,
\textit{cis}-clusters having all H atoms on the same side of the graphene
sheet, and \textit{trans}-clusters having the H atoms on both sides of the
graphene sheet in an alternating manner. For the trans-clusters a few larger clusters were further considered.
The computational method is described in Section \ref{comp}.
The calculated H structures and their energetics are presented in Section
\ref{results}. The discussion of the results, including observed trends in the
H binding with the size of H clusters, is given in Section \ref{disc}.
The results are summarized in Section \ref{conc}.
\section{Computational details}\label{comp}
The DFT calculations were performed with the plane wave based DACAPO program
package \cite{Hammer1999,Bahn2002}, applying ultra-soft pseudopotentials
\cite{Vanderbilt1990,Laasonen1993} to describe electron-ion interactions,
and the Perdew Wang functional (PW91) for the electronic exchange correlation
effects. The electron wave functions and augmented electron density were
expanded in plane waves with cutoff energies of 25 Ry and 140 Ry,
respectively. The H cluster configurations were calculated by modelling the
graphite surface with  rhombohedral, periodically repeated slabs consisting of
one graphene sheet having a 6$\times$6 surface unit cell with 72 carbon atoms,
and separated by 15 {\AA} of vacuum.
The Chadi-Cohen scheme  \cite{Chadi1973} with six special points was used  for
sampling of the surface Brillouin zone.
Binding energies of the $n$H clusters are reported per cluster using the clean graphene sheet plus $n$ separate H atoms as the reference system, i.e.:
\begin{equation}
E_b= [E(gra) + n E(\mathrm{H})] -  E(n\mathrm{H}/gra),
\end{equation}
where  $E(n\mathrm{H}/gra)$, $E(gra)$ and  $E(\mathrm{H})$ are total energies of the $n$H clusters 
at graphene, the clean graphene sheet and the free H atom, respectively.
The binding energies are well converged with respect to the number of
{\bf k}-points and the surface cell size as evidenced by
Table \ref{tab1} presenting the binding energies of the most favorable
$n$H cis-clusters ($n \le 6$) on graphene that we have encountered in
the present investigation (i.e. the monomer, the ortho-dimer ``O'',
and trimers up to hexamers to be defined below).
The H binding energy in gas-phase H$_2$ molecule calculated using
computational method described above is 2.29 eV/H. \\
\begin{table}[h]
\begin{tabular}{c|c|cccccc}
\hline
 & &\multicolumn{2}{c}{6$\times$6 cell} & {\phantom{0}}
 &\multicolumn{2}{c}{7$\times$7 cell} & \\
 Config. & \# of H atoms &\multicolumn{2}{c}{\# of {\bf k}-points} & {}
&\multicolumn{2}{c}{\# of {\bf k}-points} & \\
 & & 6 & 18 &  & 6 & 18  & \\  \hline
% \colrule
monomer & 1 & {\bf 0.77} &  0.81 &  & 0.78 & 0.82 &\\
O  & 2 & {\bf 2.73} &  2.76 &  & 2.73 & 2.75 &\\
H$_3$$^{\rm I}$ & 3 & {\bf 4.16} &  4.22 &  & 4.21 & 4.26 &\\
H$_4$$^{\rm I}$ & 4 & {\bf 6.27} &  6.32 &  & 6.28 & 6.29 &\\
H$_5$$^{\rm I}$ & 5 & {\bf 7.66} &  7.73 &  & 7.73 & 7.79 &\\
H$_6$$^{\rm I}$ & 6 & {\bf 9.62} &  9.68 &  & 9.66 & 9.69 &\\
\hline
\end{tabular}
\caption{\label{tab1}Binding energies (in eV) of $n$H (1$\le n \le 6$)
cis-clusters adsorbed on graphene, as a function of the simulation cell size and
the number of {\bf k}-points used to sample the Brillouin zone. Results
are given for the monomer, the ortho-dimer (``O'') and the most stable
structures from Table \ref{etable}.}
\end{table}
In figures \ref{dimers}-\ref{trans-big} schematic illustrations of the H islands omitting the 
ionic relaxation patterns are show. However, all structures were fully 
relaxed using the Broyden-Fletcher-Goldfarb-Shanno algorithm \cite{bfgs}. 
The activation energies for diffusion of H atoms and for the H$_2$ recombination 
were calculated applying the nudged elastic band method \cite{neb} 
using at least seven configurations to model reaction paths.
For sequences of reaction steps, we use the term \textit{net barrier} to 
denote the energy difference between the initial state and the highest 
potential energy point along the path. This highest energy point will be 
a transition state of some $n$'th reaction step along the path and the 
net barrier becomes the sum of the reaction energy for moving from the 
first initial state to the $n$'th initial state \textit{plus} the local 
barrier in the $n$'th reaction step.

\section{Results} \label{results}
In the present paper we considered two types of adsorbed $n$H clusters:
(A) cis-clusters with all H atoms adsorbed on the same side of the graphene sheet,
and (B) trans-clusters where adsorbates bind at both sides of the graphene in an alternating 
manner. The corresponding results are presented in subsections \ref{up} and
\ref{both}.
\subsection{Cis-clusters: H adsorbed only on one side of the sheet}
\label{up}
Since the activation energy of H diffusion through graphene is higher than
4 eV \cite{Ferro2002} in the majority of experiments related to the H 
adsorption on graphite or graphene, the adsorbates will exclusively bind on 
the side of the graphene sheet exposed to the source of H atoms. Hence, 
we first consider this class of H configurations on graphene.

\subsubsection{Hydrogen monomer and short dimers}
\begin{figure}[h]
\includegraphics[width=0.8\linewidth]{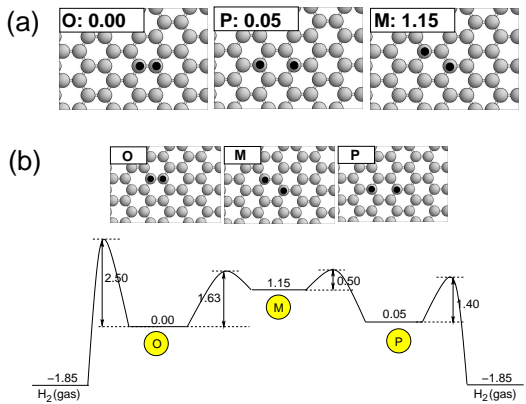}
\caption{\label{dimers} (a) Three H short-dimer configurations 
on graphene. H and C atoms are represented with small black and gray spheres, 
respectively. Energies are given relative to the energy of the O-dimer structure; 
(b) the path and potential energy diagram for H$_2$
recombination from the O-dimer configuration. The transformation from the
O-dimer to the P-dimer has a net barrier of 1.65 eV (the sum of 1.15 and 0.50 eV). 
All energies are in eV.
}
\end{figure}

The structure and binding energies of the H monomer and dimers have been studied
by several groups \cite{Sha2002,Zecho2002b,Ferro2003,Miura2003,Hornekaer2006,
Casolo2008,Zeljko2009}. Our recent publication \cite{Zeljko2009} contains
well tested results for the H monomer and a comprehensive list of hydrogen dimer
structures on graphene. The study includes H binding energies and activation
energies of the most important kinetic processes. Since the results for bigger
clusters presented in this paper will be compared to those of monomers and
short dimers, for the sake of convenience we include in Fig.\ \ref{dimers} 
and Table \ref{mono_di} relevant results of Ref. \cite{Zeljko2009}, 
calculated with the same computational set-up as used in the present study.
\begin{table}[t]
\begin{tabular}{cc|cc}
\hline
 \multicolumn{2}{c}{monomer} &\multicolumn{2}{c}{dimers} \\ \hline
 config. & binding &  config. & binding \\
 & energy  & & energy \\ \hline
 monomer &  0.77 & O   & 2.73 \\
 &  &              P   & 2.68 \\
 &  &              M   & 1.58 \\
\hline
\end{tabular}
\caption{\label{mono_di} Binding energies (in eV) of the monomer and
short dimers at graphene \cite{Zeljko2009}. 
%
%O is the ortho-dimer where the
%H atoms are on neighboring C atoms. P is the para-dimer where the H atoms
%are on C atoms that are opposite within one aromatic ring. M is the
%meta-dimer for which the H atoms are on next-nearest neighboring sites.
}
\end{table}
\subsubsection{Hydrogen trimers}
\begin{figure}[h]
\includegraphics[width=1.0\linewidth]{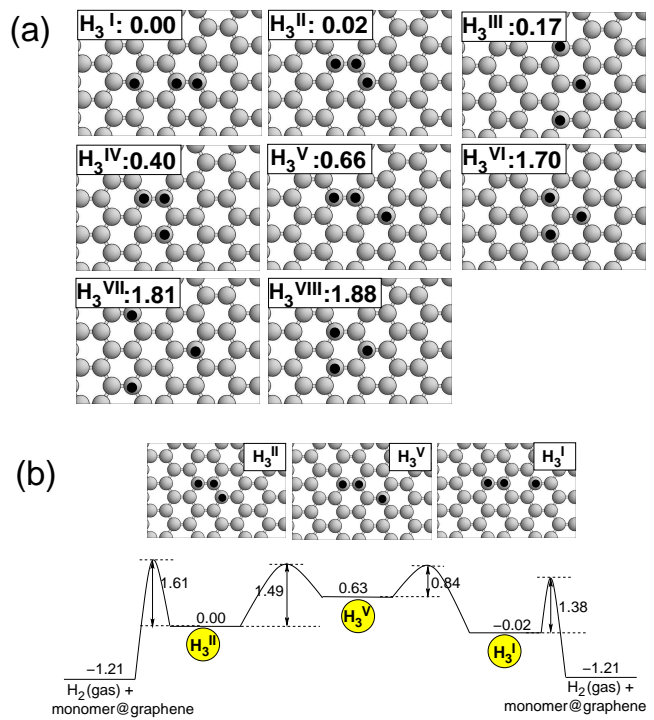}
\caption{\label{trimers} (a) H trimer configurations on graphene. Energies are
given relative to the energy of the  H$_3$$^{\rm I}$ structure; (b) the paths and potential
energy profiles for H$_2$
recombination from the  H$_3$$^{\rm I}$ and  H$_3$$^{\rm II}$ configurations. All energies are in eV.
}
\end{figure}

The trimer structures investigated in the present work are depicted in
Figure \ref{trimers}(a), together with the total energies calculated relative
to the energy of the  H$_3$$^{\rm I}$ structure, identified as the most stable among
the trimers considered. The binding energies of the H trimers are
given in Table \ref{etable}.
\begin{table*}[t]
\begin{tabular}{cc|cc|cc|cc}
\hline
 \multicolumn{2}{c}{trimers} &\multicolumn{2}{c}{tetramers} &
 \multicolumn{2}{c}{pentamers} &\multicolumn{2}{c}{hexamers}\\ \hline
 config. & binding &  config. & binding & config. & binding  & config.
 & binding\\
 & energy  & & energy  & & energy  & & energy \\ \hline
 H$_3$$^{\rm I}$    &  4.16 & H$_4$$^{\rm I}$   & 6.27 &  H$_5$$^{\rm I}$   & 7.66 &  H$_6$$^{\rm I}$  & 9.62 \\
 H$_3$$^{\rm II}$   &  4.14 & H$_4$$^{\rm II}$   & 6.10 & H$_5$$^{\rm II}$   & 7.54 &  H$_6$$^{\rm II}$ & 9.57 \\
 H$_3$$^{\rm III}$  &  3.99 & H$_4$$^{\rm III}$  & 6.06 &  H$_5$$^{\rm III}$  & 7.51 &  H$_6$$^{\rm III}$   & 9.55 \\
 H$_3$$^{\rm IV}$   &  3.76 & H$_4$$^{\rm IV}$   & 6.05 & H$_5$$^{\rm IV}$   & 7.50 &  H$_6$$^{\rm IV}$  & 9.42 \\
 H$_3$$^{\rm V}$    &  3.50 & H$_4$$^{\rm V}$   & 5.94 & H$_5$$^{\rm V}$    & 7.50 &  H$_6$$^{\rm V}$  & 9.31 \\
 H$_3$$^{\rm VI}$   &  2.46 & H$_4$$^{\rm VI}$   & 5.87 & H$_5$$^{\rm VI}$   & 7.49 &  H$_6$$^{\rm VI}$  & 9.27 \\
 H$_3$$^{\rm VII}$  &  2.35 & H$_4$$^{\rm VII}$  & 5.85 & H$_5$$^{\rm VII}$   & 7.47 &  H$_6$$^{\rm VII}$  & 9.23 \\
 H$_3$$^{\rm VIII}$ &  2.28 & H$_4$$^{\rm VIII}$  & 5.51 & H$_5$$^{\rm VIII}$  & 7.31 &  H$_6$$^{\rm VIII}$  & 9.21 \\
                  &       & H$_4$$^{\rm IX}$   & 5.42 & H$_5$$^{\rm IX}$    & 7.26 &  H$_6$$^{\rm IX}$  & 9.14 \\
                  &       & H$_4$$^{\rm X}$   & 5.29 & H$_5$$^{\rm X}$     & 7.25 &     & \\
                  &       & H$_4$$^{\rm XI}$  & 5.20 & H$_5$$^{\rm XI}$    & 7.24 &     & \\
                  &       &  H$_4$$^{\rm XII}$ & 5.12 & H$_5$$^{\rm XII}$   & 7.15 &     & \\
                  &       &                  &      & H$_5$$^{\rm XIII}$  & 7.11 &     & \\
                  &       &                  &      & H$_5$$^{\rm XIV}$   & 7.09 &     & \\
\hline
\end{tabular}
\caption{\label{etable} Binding energies (in eV) of $n$H cis-clusters at
graphene. The cluster configurations are depicted in Figs. \ref{trimers}-
\ref{hexamers}.}
\end{table*}
It is instructive to interpret the three most stable trimers (i.e.  H$_3$$^{\rm I}$,  H$_3$$^{\rm II}$ and  H$_3$$^{\rm III}$)
as being built from an ortho- or para-dimer (O- or P-dimer) with an extra atom added in an O- or a P-position.
Adding the extra atom a in meta-position (M-position) results in less stable structures
(i.e.  H$_3$$^{\rm IV}$ and  H$_3$$^{\rm V}$). Thus the presence of an
embedded M-dimer, i.e.\ involving somewhere in the H clusters two H atoms in M-position with respect to each other, 
appears to be strongly disfavored. The least stable trimers considered are triangular structures with
the atoms in M- or extended dimer-positions (i.e. H$_3$$^{\rm VI}$, H$_3$$^{\rm VII}$ and H$_3$$^{\rm VIII}$).

The  H$_3$$^{\rm I}$ and  H$_3$$^{\rm III}$ structures were also presented by Casolo {\it et al.} 
\cite{Casolo2008}, with the binding energies in very good agreement with 
the values reported here. The  H$_3$$^{\rm VII}$ trimer was suggested by Khazaei {\it et al.} 
\cite{ Khazaei2009} as a model for experimentally observed star like STM 
patterns, but appears as one of the least favorable trimer structures according
to our work. Since we further calculate an energy barrier for H diffusion to neighbouring C sites
of less than 0.65 eV for the  H$_3$$^{\rm VII}$ configuration, it is unlikely to be stable at 
temperatures up to 600 K \cite{Hornekaer2007}. A very comprehensive list of 
trimer structures is reported by Roman {\it at al.} \cite{Roman2009}. The most 
favorable structures ( H$_3$$^{\rm I}$,  H$_3$$^{\rm II}$ and  H$_3$$^{\rm III}$) observed in the present work are in full 
agreement with the results in Ref. \cite{Roman2009}. \\
Considering the associative desorption of H$_2$ from the trimer structures
we find that it occurs via an embedded P-dimer. The calculated pathways are
sketched in Fig.\ \ref{trimers}(b) for the two most favorable structures (i.e.  H$_3$$^{\rm I}$ and  H$_3$$^{\rm II}$). For configuration  H$_3$$^{\rm I}$ that already contains an embedded P-dimer, the
barrier is calculated to 1.38 eV, which is close to the one found for the isolated P-dimer (1.40 eV\cite{Hornekaer2006,Zeljko2009} -- see Fig.\ \ref{dimers}(b)).
At the transition state along this pathway the H$_2$ molecule is nearly 
parallel to the surface with the H-H bond-length of $\sim$1.2 {\AA}. Very 
similar geometries of transition states are accounted for other reaction paths 
shown in Figs. \ref{tetramers} and \ref{hexamers}. 
The  H$_3$$^{\rm II}$ trimer does not contain an embedded P-dimer and the associative desorption from
the embedded O-dimer is calculated to be associated with a barrier of 1.61 eV.
An alternative scenario includes several H diffusion steps transforming the  H$_3$$^{\rm II}$ trimer via the  H$_3$$^{\rm V}$ trimer to the  H$_3$$^{\rm I}$ trimer for which the associative desorption
can occur. Along this path, the highest activation energy becomes that of the
transition from the  H$_3$$^{\rm II}$ to the  H$_3$$^{\rm V}$ trimer, which is 1.49 eV, see Fig.\ \ref{trimers}(b).
\subsubsection{Hydrogen tetramers}
\begin{figure}
\includegraphics[width=1.0\linewidth]{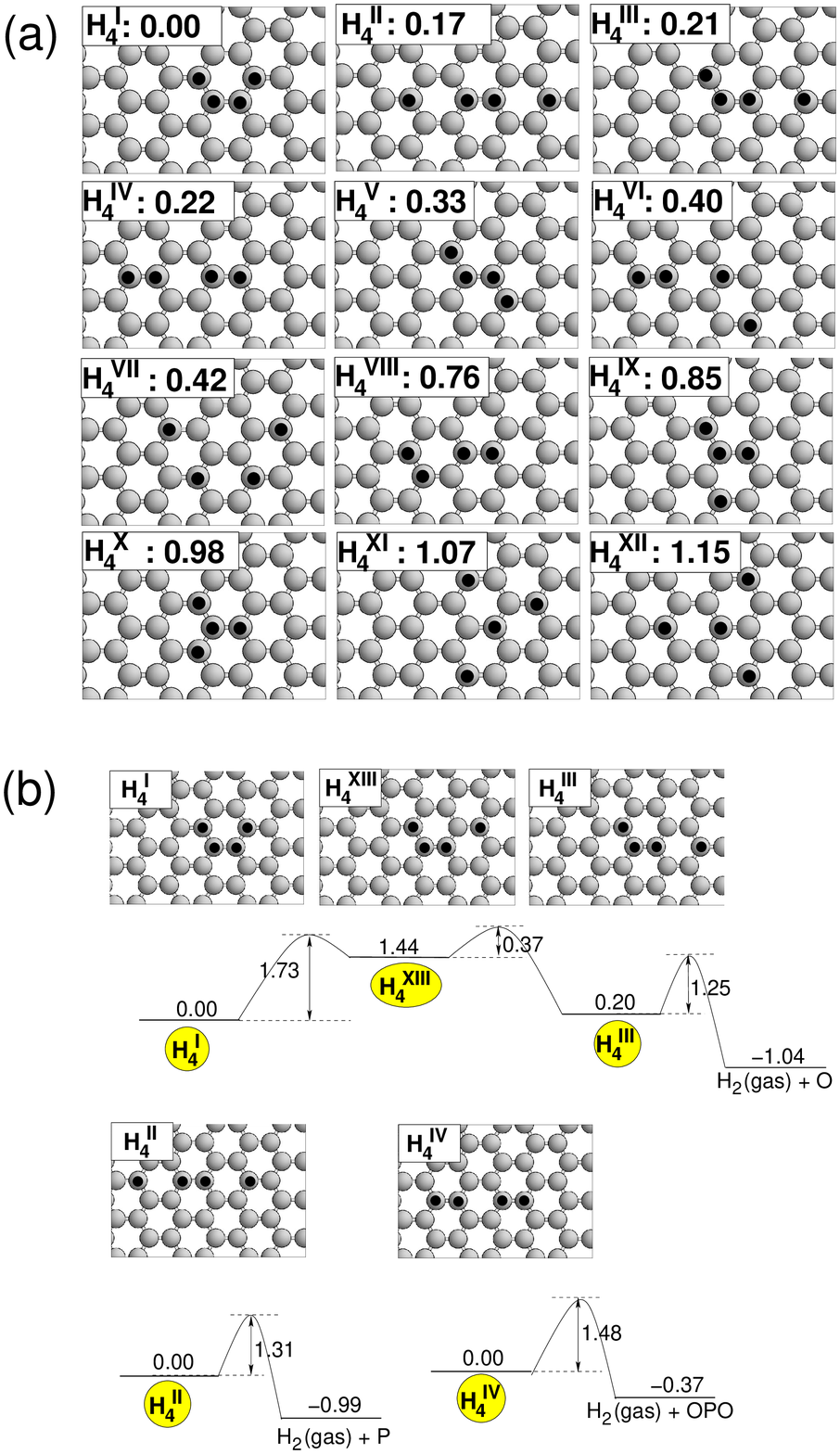}
\caption{\label{tetramers} (a) H tetramer configurations on graphene. Potential energies
are given relative to the energy of the  H$_4$$^{\rm I}$ structure.
(b) The pathways for H$_2$ recombination from the  H$_4$$^{\rm I}$,  H$_4$$^{\rm II}$,  H$_4$$^{\rm III}$ and  H$_4$$^{\rm IV}$ configurations. The OPO is one of the long-dimer configurations from Ref. \cite{Zeljko2009}. All energies are in eV.
}
\end{figure}
The most favorable tetramer structures found are illustrated in Figure
\ref{tetramers}(a) and their binding energies are given in Table \ref{etable}. There are many low energy tetramers and a common feature
for the very most stable of them (i.e. H$_4$$^{\rm I-VII}$) seems to be that they are composed of one
of the most stable trimers with an extra atom attached in an O- or a P-position.
It is interesting to note that a triangular arrangement of the atoms seems to be particularly unfavorable. 
Comparing H$_4$$^{\rm XI}$ and H$_4$$^{\rm XII}$ one might guess based on the previous discussion that H$_4$$^{\rm XII}$ would be the more stable, 
since it contains exclusively atoms in P-positions, whereas in H$_4$$^{\rm XI}$ one atom is moved to a M-position. 
%
% Feb2011
%
%Since this 
%is not the case, we deduce that the triangular symmetry itself is unfavorable. 
The same arguments apply when comparing H$_4$$^{\rm IX}$ and H$_4$$^{\rm X}$. 
These triangular structures are unfavorable due to three H atoms  
adsorbed on the same C sublattice. Such imbalance in the occupation of two
carbon sublattices creates additional unpaired electrons in the graphene lattice, 
increasesing the total energy of the system.
Note that for the trimers considered triangular structures were also very 
unfavorable.
%
%In order to save computing time, for unfavorable configurations with
%the total energy more than 1 eV higher than in the configuration C, the
%calculations are restricted to the 5$\times$5 and 6$\times$6  unit cells.
%
The pathways for H$_2$ recombination from the four most favorable
configurations are depicted in Figure \ref{tetramers}(b). As for the
trimers, the recombination occurs from an embedded P-dimer, either
directly [for the  H$_4$$^{\rm II}$,  H$_4$$^{\rm III}$ and  H$_4$$^{\rm IV}$ tetramers], or via H diffusion from the
 H$_4$$^{\rm I}$ over the  H$_4$$^{\rm XIII}$ to the  H$_4$$^{\rm III}$ structure, followed by H$_2$ associative
desorption, as shown in Figure \ref{tetramers}(b).
The net barrier for H$_2$ formation from the  H$_4$$^{\rm I}$ tetramer becomes
1.81 eV (the sum of 1.44 eV and 0.37 eV), which demonstrates a high
stability of this configuration.
Other possible scenarios for H$_2$ recombination from tetramers
are discussed in Ref. \cite{Luntz2009}.
\subsubsection{Hydrogen pentamers}
\begin{figure}
\includegraphics[width=1.0\linewidth]{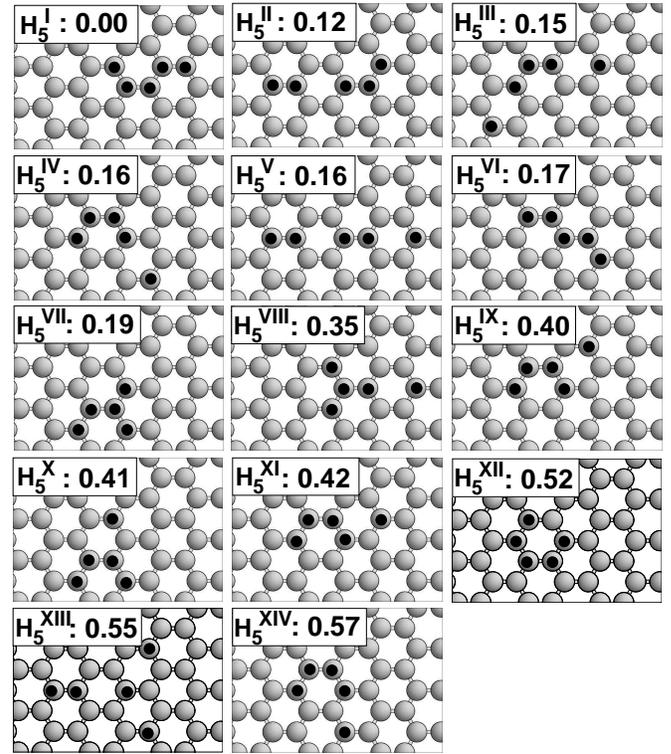}
\caption{\label{pentamers} H pentamer configurations on graphene. Potential
energies (in eV) are given relative to the energy of the  H$_5$$^{\rm I}$ structure.
}
\end{figure}

The most stable pentamer structures in Figure \ref{pentamers} are formed by
attachment of an additional H atom in an O- or a P-position to one of the preferential tetramer
structures. The binding energies of the configurations considered are provided in
Table \ref{etable}.
The binding energies of the seven most stable structures are within
an energy window of 0.2 eV. Although the other studied configurations are
less favorable, the energy difference between the least (the  H$_5$$^{\rm XIV}$) and the
overall most stable structure (the  H$_5$$^{\rm I}$) is smaller than 0.6 eV. \\ 
%
%We again note that structures containing motifs of triangular or distorted triangular shape 
%(i.e. H$_5$$^{\rm VIII}$ and H$_5$$^{\rm XIII}$) are among the least stable ones.\\
%
The activation energies for associative desorption are not calculated, since
we expect that the H$_2$ formation occurs along pathways that are qualitatively
similar to those of the tetramers, i.e.\ via direct associative recombination from
embedded P-dimers or several steps of H diffusion between adjacent C sites, prior to the recombination from a P-dimer state.
\subsubsection{Hydrogen hexamers}
\begin{figure}
\includegraphics[width=1.0\linewidth]{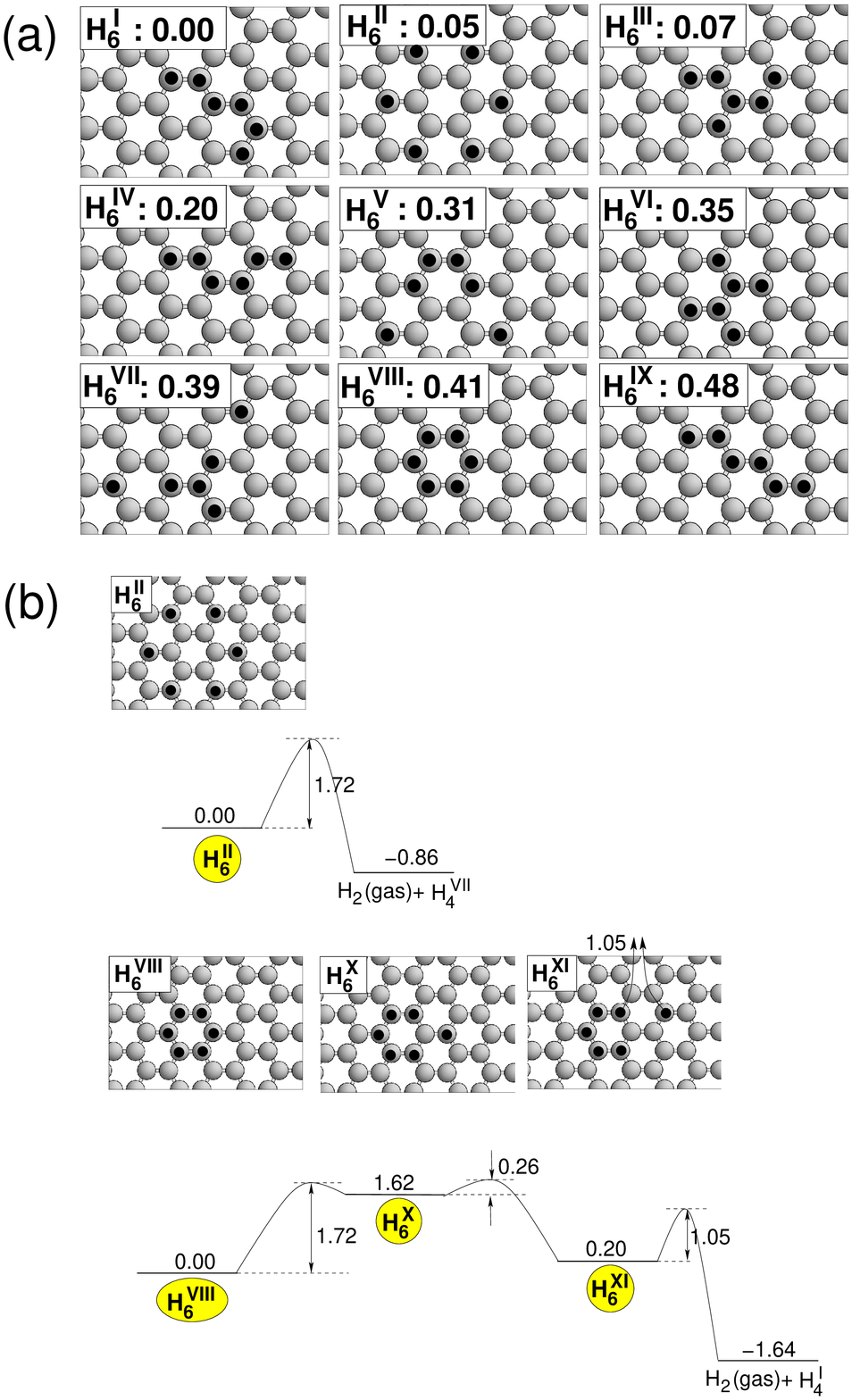}
\caption{\label{hexamers} (a) H hexamer configurations on graphene. Potential
energies are given relative to the energy of  H$_6$$^{\rm I}$ structure. (b) The reaction 
paths for H$_2$ recombination from the  H$_6$$^{\rm II}$ and  H$_6$$^{\rm VIII}$ configurations. All energies 
are in eV.}
\end{figure}

Hexamers are the largest H clusters considered systematically in the present study.
In Fig.\ \ref{hexamers}(a) the most stable structures identified are shown.
The most favorable structure,  H$_6$$^{\rm I}$, has a total
binding energy of 9.62 eV. 
Following the general trend outlined, the most stable hexamers contain motifs from stable smaller clusters and have all atoms in O- or P-positions. 
It is interesting to note that the  H$_6$$^{\rm I}$ and the  H$_6$$^{\rm II}$
structures have almost identical stability despite being quite different in
geometry; the  H$_6$$^{\rm I}$ is of low symmetry and has H atoms in O-positions while the  H$_6$$^{\rm II}$ is
of high symmetry and has H atoms in P-positions. Generally, very many hexamer structures are found that differ 
only by a few tenths of an eV.
For most of the structures presented in Fig.\ \ref{hexamers}(a) the
mechanism for the H$_2$ recombination and corresponding energy
barriers are expected to be very similar to those determined for the
dimers, trimers and tetramers. Thus, we did not perform the
corresponding calculations.  The reaction paths were however carefully
calculated for the  H$_6$$^{\rm II}$ and  H$_6$$^{\rm VIII}$ structures, since both configurations
possess ideal hexagonal symmetry.  The breaking of the high symmetry
of these two configurations might lead to activation energies for
H$_2$ recombination that are higher than the typical values found for
other H structures. The results for the H$_2$ formation from the  H$_6$$^{\rm II}$
and  H$_6$$^{\rm VIII}$ hexamers are presented in Fig.\ \ref{hexamers}(b). Indeed, the
activation energies of 1.72 eV and 1.88 eV, calculated for  H$_6$$^{\rm II}$ and  H$_6$$^{\rm VIII}$,
respectively, indicate high kinetic stability of these structures with 
respect to molecular hydrogen formation. 
Ferro et al. \cite{Ferro2009} have suggested the  H$_6$$^{\rm II}$ hexamer
structure as a possible candidate for the experimentally observed very stable
starlike STM patterns \cite{Hornekaer2007}.

\subsection{Trans-clusters: H adsorbed on both sides of the sheet}
\label{both}
\begin{figure}
\includegraphics[width=1.0\linewidth]{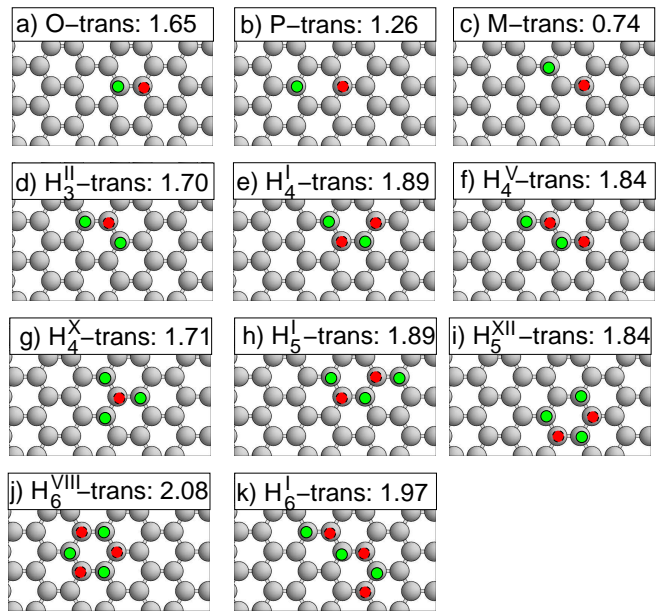}
\caption{\label{opposite} H structures with atoms adsorbed on both sides of
the graphene sheet: (a)-(c) trans-dimers; (d) trans-trimer; (e)-(g) trans-tetramers;
(h)-(i) trans-pentamers; (j)-(k) trans-hexamers. Adatoms from opposite sides of the sheet
are colored differently. Binding energies per H atom are given in eV.
}
\end{figure}

A common feature of the dimers in Ref.\cite{Zeljko2009} and the cis-clusters
presented in Figures \ref{trimers} to \ref{hexamers} is that
all H atoms are adsorbed on the same side of the graphene sheet. Given a very
high energy barrier for H diffusion through the graphene layer this scenario
is most likely to occur in defect-free samples deposited at crystalline 
surfaces. Yet, in free-standing graphene, graphene samples with defects or in 
small graphene patches, diffusion barriers at the defects or
edges could be significantly lower than at the perfect sheet,
opening routes for H adsorption on both sides of the layer. It turns out that
such clusters -- trans-clusters -- are much more stable than those considered in the previous
section. In the following, we present the results of our systematic investigation
of trans-clusters with 2-6 H atoms, showing, however, 
only the very most stable structures identified. We further selectively
chose to study a few larger trans-clusters that will be presented and discussed.\\
The O-trans-dimer formed by adsorption of two H atoms in the configuration
depicted in Fig.\ \ref{opposite}a is very stable with a total binding
energy of 3.30 eV. This is by 0.54 eV higher than the binding in the
O-dimer with the H atoms on the same side of the graphene sheet, i.e.\ the O-cis-dimer.
The high stability of the O-trans-dimer configuration has already been
reported by Roman {\it et al.} \cite{Roman2009b}.
The high stability of the O-trans-dimer appears to be related to
the immediate proximity of the C atoms, since expanding the
trans-dimer forming the \textit{P-trans-} and \textit{M-trans-dimers}
of Fig.\ \ref{opposite}b and Fig.\ \ref{opposite}c leads to binding
energies that are smaller by 0.79 and 1.82 eV, respectively, than
that of the O-trans-dimer. In view of this finding, the search for the structures of the most stable larger H clusters was limited to
structures in which all neighboring H atoms were configured as
in the O-trans-dimer.\\
\begin{table}[h]
\begin{tabular}{ccccccc}
\hline
\multicolumn{2}{c}{configuration} &\phantom{000} &\# of H atoms&\phantom{000} & \multicolumn{2}{c}{bind. energy}\\ 
& & & & & (eV)&(eV/H)  \\
O-trans   & [Fig. \ref{opposite}(a)]& &2 & &3.30 &1.65 \\
P-trans   & [Fig. \ref{opposite}(b)]& &2 & &2.51 &1.26 \\
M-trans   & [Fig. \ref{opposite}(c)]& &2 & &1.48 &0.74 \\ 
 H$_3$$^{\rm II}$-trans  & [Fig. \ref{opposite}(d)]& &3 & &5.11 &1.70 \\
 H$_4$$^{\rm I}$-trans  & [Fig. \ref{opposite}(e)]& &4 & &7.56 &1.89 \\
 H$_4$$^{\rm V}$-trans  & [Fig. \ref{opposite}(f)]& &4 & &7.36 &1.84 \\
 H$_4$$^{\rm X}$-trans & [Fig. \ref{opposite}(g)]& &4 & &6.83 &1.71 \\
 H$_5$$^{\rm I}$-trans  & [Fig. \ref{opposite}(h)]& &5 & &9.44 &1.89 \\
 H$_5$$^{\rm XII}$-trans & [Fig. \ref{opposite}(i)]& &5 & &9.20 &1.84 \\
 H$_6$$^{\rm VIII}$-trans  & [Fig. \ref{opposite}(j)]& &6 & &12.47 &2.08 \\
 H$_6$$^{\rm I}$-trans  & [Fig. \ref{opposite}(k)]   & &6 & &11.81 &1.97 \\
%Tr10      & [Fig. \ref{trans-big}]  & &10 & &1.83 \\
%Tr12      & [Fig. \ref{trans-big}]  & &12 & &2.08 \\
%Tr12a     & [Fig. \ref{trans-big}]  & &12 & &      \\
%Tr13      & [Fig. \ref{trans-big}]  & &13 & &2.17 \\
%Tr16      & [Fig. \ref{trans-big}]  & &16 & &2.22 \\
%Tr22      & [Fig. \ref{trans-big}]  & &22 & &2.23? \\
\hline
\end{tabular}
\caption{\label{clusters}Binding energies of hydrogen trans-clusters adsorbed 
on graphene.}
\end{table}
The  H$_3$$^{\rm II}$-trans-trimer structure shown in  Fig.\ \ref{opposite}d is the most favorable trans-trimer 
structure found in our study. The binding energy of 5.11 eV,
is by 0.89 higher than in the  H$_3$$^{\rm I}$ cis-trimer. \\
Two almost equally stable trans-tetramers (H$_4$$^{\rm I}$-trans and H$_4$$^{\rm V}$-trans) were found as shown in Fig.\ \ref{opposite}e-f with binding energies of 7.56 and 7.36 eV.
The triangular shaped trans-tetramer in Fig.\ \ref{opposite}g is slightly less stable with a binding energy of 6.83 eV, which is 0.73 eV smaller than Fig.\ \ref{opposite}e.
This indicates that also for trans-clusters a triangular arrangement of the H atoms is disfavored,
and can in the same way as for the triangular shaped cis-clusters be related to an imbalance in the 
number of H atoms adsorbed on the two C sublattices.
The most stable configuration, Fig.\ \ref{opposite}e, is as much as 1.26 eV more
stable than the H$_4$$^{\rm I}$-cis-tetramer in Fig.\ \ref{tetramers}.\\
Two H trans-pentamers (H$_5$$^{\rm I}$-trans and H$_5$$^{\rm XII}$-trans) were investigated as shown in Fig.\ref{opposite}h
and \ref{opposite}i. The calculated total H binding energies were 9.44 and
9.20  eV, respectively. The structure in Fig.\ \ref{opposite}h is by 1.71 eV
more stable than the H$_5$$^{\rm I}$-cis pentamer in Fig.\ \ref{pentamers}. \\
The trans-hexamer in  Fig.\ \ref{opposite}j (H$_6$$^{\rm VIII}$-trans) is a particularly stable
structure with a total binding energy
of 12.47 eV which is 0.66 eV larger than that of the configuration in
Fig.\ \ref{opposite}k (H$_6$$^{\rm I}$-trans), and as much as 2.79 higher than the binding of the
most stable cis-structure, H$_6$$^{\rm I}$, in Fig.\ \ref{hexamers}. \\
\begin{figure}
\includegraphics[width=1.0\linewidth]{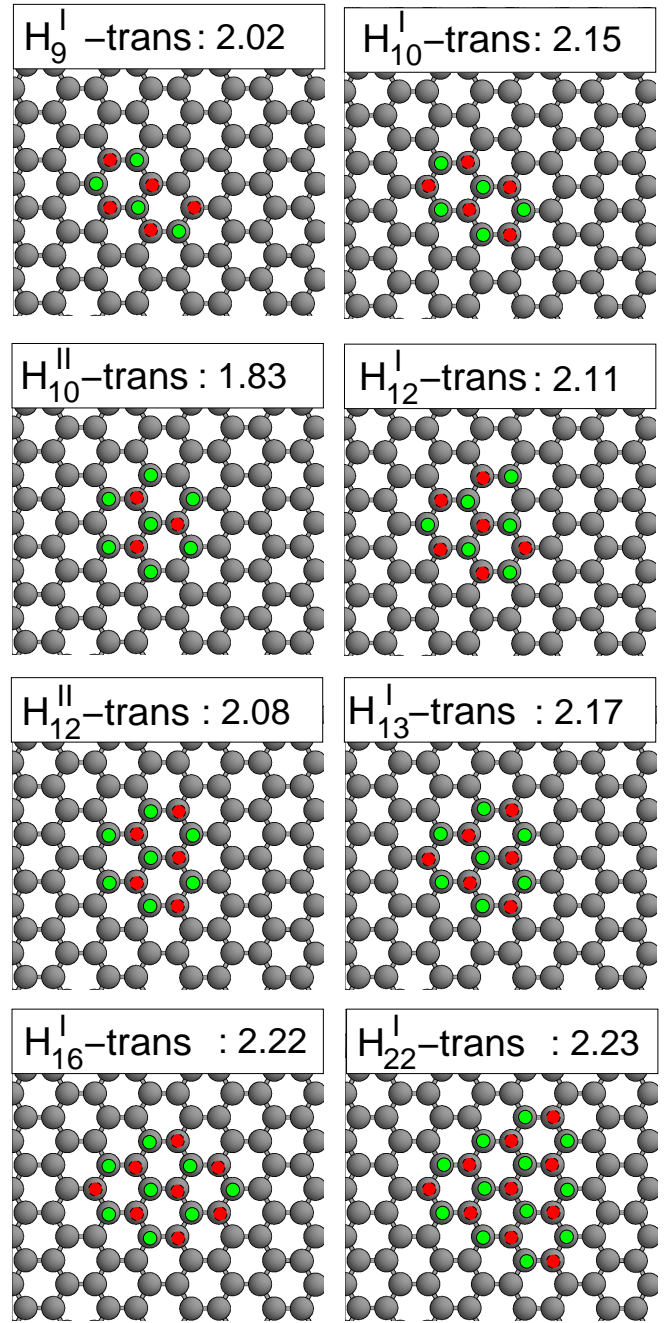}
\caption{\label{trans-big} Bigger H structures with atoms adsorbed on both 
sides of the graphene sheet. Adatoms from opposite sides of the sheet are 
colored differently. Binding energies per H atom are given in eV. 
}
\end{figure}
The graphane, fully hydrogenated graphene with hydrogen atoms adsorbed at carbon atoms on both sides of the sheet in an alternating 
manner \cite{Sofo2007,Elias2009}, can be considered as the infinite trans-cluster. The H binding energy 
in graphane is 2.49 eV per atom, which is considerably more than the 2.08 eV 
calculated for the  H$_6$$^{\rm VIII}$-trans hexamer. Thus, with an increase in the size of 
trans-clusters we expect higher H binding energies, which should approach
the value calculated for graphane. To examine this expected trend we considered
several bigger trans-clusters. The larger trans-clusters considered are shown in
Fig. \ref{trans-big}. They have all been constructed as truncated pieces of
graphane embedded in graphene. The most stable of the selected clusters are 
exclusively composed of closed hydrogenated carbon hexagons, already identified
as the most stable small trans-clusters (H$_6$$^{\rm VIII}$-trans). 
To further study the stability of these structures we included in our investigation 
configurations with  missing H atoms on carbon hexagons at the cluster edges, 
which results in the occurrence of embedded M-dimers.
Apart from the clusters having various sizes the primary difference is the configuration 
of the edge atoms. The H binding energies quoted in the figure do indeed increase with cluster size. This will be analyzed further in the Discussion Section.

\subsection{Hydrogen induced magnetism}
In a bipartite lattice, any imbalance in number of sites belonging
to each of two sublattices leads to a magnetic ground state
\cite{Lieb1989,Casolo2008,Verges2010}. One of the possible mechanisms to induce
such an imbalance on graphene is adsorption of H atoms as has been discussed in the recent
literature \cite{Oleg2007,Ferro2008,Casolo2008,Lei2008}.
According to our calculations, upon adsorption of a single H atom on graphene,
small magnetic moments are observed at several C atoms
in its vicinity. The calculated magnetic moments at the individual atoms are
smaller than 0.1 $\mu_{\rm B}$. Concerning H dimers, non-zero spin density
is observed only for structures with both hydrogens adsorbed on the C atoms
from the same sublattice.
According to the DFT calculations, these structures are unfavorable,
and therefore not expected to form in experiments. The energy gain due to
spin-polarization does not affect the relative stability of the most favorable
dimer structures obtained from non-spin-polarized calculations.
This has already been demonstrated in our previous publication 
\cite{Zeljko2009}. A similar effect of the spin-polarization is found in the present work for the energetics 
of the investigated H cis configurations with three, four, five and six atoms. 
However, several favorable trans-clusters investigated here (H$_4$$^{\rm X}$-trans, H$_{12}$$^{\rm II}$-trans, and H$_{22}$$^{\rm I}$-trans)  
carry a total magnetic moment of 2 $\mu_{\rm B}$. We note that for these clusters,
there is an imbalance between the number of H atoms on the two sides of the
graphene sheet and that such graphane islands embedded in the graphene lend themselves as possible building blocks for graphene-based magnetic materials.\\
\section{Discussion}\label{disc}
\begin{figure}
\includegraphics[width=0.6\linewidth]{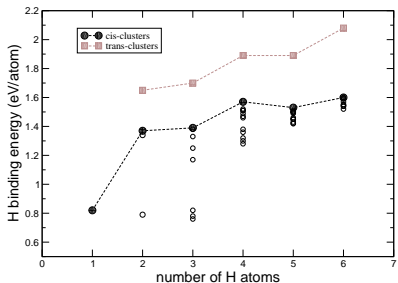}
\caption{\label{binding} The variation of the H binding energy with the size of
cis- (circles) and trans-clusters (squares). The values corresponding to the most stable cis-structures (O-dimer, H$_3$$^{\rm I}$-trimer, 
H$_4$$^{\rm I}$-tetramer, H$_5$$^{\rm I}$-pentamer and H$_6$$^{\rm I}$-hexamer) are shown as big circles. Results for less stable cis-clusters 
from Tables \ref{mono_di} and \ref{etable} are given as small open circles. 
}
\end{figure}

%
% Trends in binding energies
%
The H adsorption on graphene in compact cis-cluster structures, described in
Sec. \ref{results}, is thermodynamically  preferential compared to the
adsorption of isolated H  atoms. The binding energy of the H monomers is
0.77 eV.  Already for the smallest clusters, such as O-dimers and H$_3$$^{\rm I}$ trimers
the binding per H atom increases to 1.37 and 1.39 eV, respectively.
The H binding in H$_4$$^{\rm I}$ tetramers is 1.57 eV, and slightly smaller in H$_5$$^{\rm I}$ pentamers,
where we calculated binding energy of 1.53 eV per H atom. The strongest
binding is evaluated for H$_6$$^{\rm I}$ hexamers, with the value of 1.60 eV per H atom.
The evolution of the H binding energy with the size of the clusters 
is depicted by circles in Fig.\ \ref{binding}.\\
If we consider the binding configurations in compact trans-cluster structures
with H atoms adsorbed on both
sides of the graphene sheet, Fig.\ \ref{opposite}, the trend in calculated
values is similar: 1.65 eV (dimer), 1.7 eV (trimer), 1.89 eV (tetramer and
pentamer), and 2.08 eV (hexamer). These binding energies are depicted by squares in Fig.\ \ref{binding} and are seen to be shifted to significantly 
higher values than in the clusters with H atoms adsorbed only on one side of the graphene sheet. 
\begin{figure}
\includegraphics[width=0.5\linewidth]{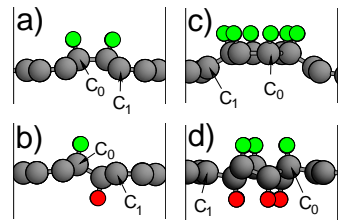}
\caption{\label{U-26} The side views of (a) O-cis-dimer; (b) O-trans-dimer;
(c) H$_6$$^{\rm VIII}$-cis-hexamer; (d) H$_6$$^{\rm VIII}$-trans-hexamer. The C atoms 
decorated with H are denoted as C$_0$ and their nearest neighbours 
without adsorbed H atoms as C$_1$. Adatoms from opposite sides of the 
sheet are colored differently.
}
\end{figure}
%
%We find that the high stability of the trans-structures is 
%due to favorable binding geometries when the adsorbed H atoms are on both 
%sides.
%
\begin{table*}[t]
\begin{tabular}{ccccccc}
\hline
% & \multicolumn{2}{c}{configuration} & & \multicolumn{2}{c}{configuration} \\
       &  O-cis  & O-trans  & & &  H8-cis  & H8-trans \\ 
$\angle$ H-C$_0$-C$_0$ &  104.84 & 106.48 & \phantom{000} &$\angle$ H-C$_0$-C$_0$ &  100.81 & 107.43 \\
$\angle$ H-C$_0$-C$_1$ &  101.75 & 106.18 & &$\angle$ H-C$_0$-C$_1$ &   98.31 & 105.54 \\
$\angle$ C$_1$-C$_0$-C$_0$ &  116.01 & 111.42& &  $\angle$ C$_0$-C$_0$-C$_0$ &  120.00 & 108.47 \\
$\angle$ C$_1$-C$_0$-C$_1$ &  113.68 & 114.56& & $\angle$ C$_1$-C$_0$-C$_0$ &  115.51 & 113.77 \\
\end{tabular}
\caption {\label{angles} The angles (in deg.) calculated for dimers (O-cis and
O-trans) and hexamers (H$_6$$^{\rm VIII}$-cis and (H$_6$$^{\rm VIII}$-trans) in Fig. \ref{U-26}.}
\end{table*}
Although our geometry for the O-trans dimer is slightly different from 
the one found by  Boukhvalov {\it et al.} \cite{Boukhvalov2008},
their explanation of particularly favorable H adsorption on 
graphene observed in O-trans dimers is fully applicable 
to the clusters considered in this study. In Fig.\ \ref{U-26}a-b side views of
the O-cis- and O-trans-dimers
%and the H8-cis- and H8-trans-hexamers
are shown and the carbon atoms are labeled. Using these labels, the
H-C-C and C-C-C angles in the structures are given in Table
\ref{angles}. From the Table it is seen that the angles calculated for trans-clusters are 
closer to the ideal tetrahedron value of 109.5$^\circ$  than those resulting for
the cis-structures. Thus, in trans-clusters the H binding configurations allow 
creation of nearly perfect tetrahedral surroundings of C atoms directly 
involved in the interaction with adsorbates which results in an additional gain in
the H chemisorption energy.
%The side views of the \textit{O-cis-dimer} and \textit{O-trans-dimer} 
%structures are depicted in Fig.\ \ref{U-26}a-b. 
%
%In both dimers, a single C=C double bond is opened and $sp^3$ bonding is
%attained locally, but the high stability of the \textit{O-trans-dimer}
%over the \textit{O-cis-dimer} suggests that a less costly strain field
%is set up in the former, where the two C atoms involved relax in
%opposite directions away from the plane of the graphene sheet as
%opposed to the latter, where the C atoms relax in the same direction.
%
An effect similar to that identified for the
O-dimers is found when comparing the (H$_6$$^{\rm VIII}$
cis-hexamer [Fig.\ \ref{hexamers} and Fig.\ \ref{U-26}c] and the 
corresponding trans-hexamer structure [Fig.\ \ref{opposite}j and Fig.\ 
\ref{U-26}d] where all six C atoms from the hexagon decorated with H adatoms 
share favorable tetrahedral coordination. 
For bigger trans-clusters with 
fully hydrogenated carbon rings [H$_{10}$$^{\rm I}$, H$_{13}$$^{\rm I}$, H$_{16}$$^{\rm I}$ and H$_{22}$$^{\rm I}$ in Fig. \ref{trans-big}] the H
binding is further enhanced, slowly approaching the value encountered in 
graphane.\\  
A similar investigation has not been carried out for cis-clusters since for
this class of H adsorption configurations trends are less obvious.
A full monolayer of H atoms can not be adsorbed on the same side of 
the graphene sheet. The maximum H coverage on graphene resulting in one or 
several stable adsorption configurations is also unknown. Thus, the trends in the 
structure and stability of bigger H cis-clusters on graphene is an open issue 
out of the scope of the present study. \\

\subsection{Trends in stability of large trans-clusters}
\begin{figure}[h]
\includegraphics[width=0.6\linewidth]{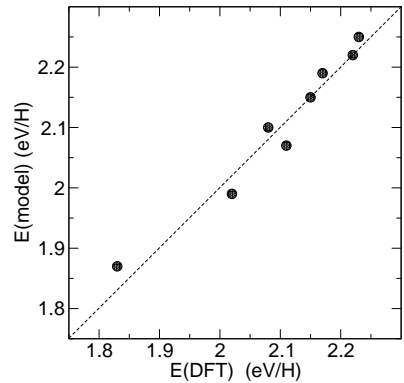}
\caption{\label{model} The correlation between hydrogen binding energies for the clusters in 
Fig. \ref{trans-big}, obtained from DFT calculations and those produced applying Eq. \ref{model_eq}.
} 
\end{figure}

%
% Simple model
%
To rationalize the DFT results obtained for the big trans-clusters, Fig. \ref{trans-big}, we construct in the following a simple model
that reproduces the trends in stability of these clusters. In the model we identify closed hydrogenated
carbon hexagons and a maximal ratio of inside atoms to edge atoms in the cluster as the most important
structural motifs causing high cluster stability. On the other hand, removing a H atom from a closed
hydrogenated carbon hexagon, i.e. introducing an embedded M-dimer, is modeled with an energy cost.
Graphane, which can be considered as an infinite trans-cluster with no edge atoms, thus represents the
highest achievable cluster stability.\\
In the model, the binding energy per H, $E(model)$, is defined as:
%follows: 
\begin{equation}
E(model) = (N_{inside} E^B_{inside} + N_{edge} E^B_{edge} - N_M E_M)/n \label{model_eq}. 
\end{equation}
where $N_{inside}$ and $N_{edge}$ are the number of inside and edge
atoms, respectively and where $N_M$ is the number of carbon sites where the
lack of an H atom causes the appearance of an embedded M-dimer. The
three energy terms, $E^B_{inside}$, $E^B_{edge}$, and $E_M$ are fixed
from calculated DFT values, i.e.\ they are not considered ajustable
parameters. For $E^B_{inside}$ we use 2.45 eV, which is the H-binding
energy of graphane when the lattice constant of graphene, 1.42 {\AA} is
used (using the self-consistent graphane lattice constant of 1.46 {\AA}
it would be 2.49 eV). For $E^B_{edge}$ we use 2.08 eV, which is the H
binding energy of $H_6^{VIII}{\rm -trans}$ that has H atoms
exclusively in edge sites. Finally, the energy penalty of introducing M-dimers,
$E_M$, we extract from the calculated DFT energy difference between
the $E_6$$^{VIII}$-trans and the $E_5$$^{XII}$-trans clusters [see Fig. \ref{opposite}], i.e. $E_M$= 5*2.08 - 5*1.84 = 1.2 eV. \\
For each of the clusters from Fig. \ref{trans-big} the relevant parameters and calculated energies  
are given in Table \ref{table_trans} and Fig. \ref{model}.
\begin{table}[h]
\begin{tabular}{c|cccccc}
\hline
 Config. &$N_{inside}$ & $N_{edge}$ &  $N_M$ & \multicolumn{2}{c}{$E(model)$}\phantom{0}& $E(DFT)$\\
         & & & &\multicolumn{2}{c}{(eV/H)} & (eV/H)\\\hline
H$_{9}$$^{\rm I}$ & 1 & 8 & 1 &&  1.99   & 2.02 \\
H$_{10}$$^{\rm I}$ & 2 & 8 & 0 &&  2.15  & 2.15 \\
H$_{10}$$^{\rm II}$ & 4 & 6 & 3 && 1.87 & 1.83 \\
H$_{12}$$^{\rm I}$  & 3 & 9 & 1 && 2.07 & 2.11 \\
H$_{12}$$^{\rm II}$ & 4 & 8 & 1 && 2.10 & 2.08 \\
H$_{13}$$^{\rm I}$ & 4 & 9 & 0 && 2.19 & 2.17 \\
H$_{16}$$^{\rm I}$ & 6 &10 & 0 && 2.22 & 2.22 \\
H$_{22}$$^{\rm I}$ &10 &12 & 0 && 2.25 & 2.23 \\\hline
\end{tabular}
\caption{\label{table_trans} The parameters and H binding energies from model, $E(model)$, and from the the full DFT calculations, $E(DFT)$.
}
\end{table}
An excellent correlation between DFT values for H binding in trans-clusters from Fig. 
\ref{trans-big} and the results obtained using Eq. \ref{model_eq} demonstrates that the general trends
in the stability of graphane-like patches on graphene can be rationalized from the  simple 
model with the parameters all determined by DFT calculations.

\subsection{Trends in H diffusion barriers in cis-clusters}
\begin{figure}
\includegraphics[width=0.6\linewidth]{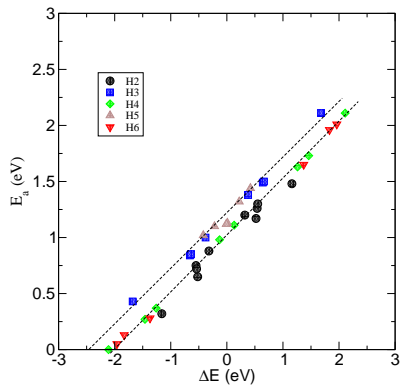}
\caption{\label{bep} Br{\o}nsted-Evans-Polanyi plot for selected H cis-clusters
from Figures \ref{trimers}-\ref{hexamers}. For each of the considered H clusters 
we calculated the energy barrier (E$_a$) for H diffusion to one of
nearest non-hydrogenated C sites. $\Delta$E is a difference in total energies 
of a cluster configuration produced upon H diffusion and the initial one.
The dashed lines represent corresponding linear relations between E$_a$ and  
$\Delta$E for structures with odd and even number of H atoms given by
Eq. \ref{eq-diff}.
}
\end{figure}

Even in the cis-structure with the highest H binding energy, i.e. the 
configuration $H_6^I$, the H binding energy is smaller than in the gas-phase 
H$_2$ molecule. Hence, all cis-clusters considered in the 
present study are metastable against associative H$_2$ desorption. Yet, at sufficiently low temperatures this process will 
be hindered due to the considerable activation energies that we calculate. According to our previous studies \cite{Hornekaer2006,
Zeljko2009} the lowest activation energy for H$_2$ recombination from H dimers
is calculated for the P-dimer state. In other dimer structures and for many of the large H clusters, the recombination
includes one or several diffusion steps of H atoms leading to the formation
of the embedded P-dimer, followed by the H$_2$ associative desorption.\\
To lay the grounds for understanding these events, we performed a
comprehensive set of calculations regarding energy barriers for H diffusion
on graphene, considering most of the structures presented in
Figs. \ref{tetramers}-\ref{hexamers}. According to the universality concept
of N{\o}rskov and co-workers \cite{universality,hammertopics}, the activation energies
for diffusion of atoms on (metal) surfaces correlate with the differences between their
binding energies in the initial and final states. Correlation diagrams--so-called
Br{\o}nsted-Evans-Polanyi (BEP) plots--reveal that weaker bound
adsorbates generally diffuse with lower activation energies. The BEP plot in
Fig.\ \ref{bep} clearly demonstrates this trend for H diffusion at graphene.
The barrier for diffusion of an isolated H atom ($\Delta$E=0) is 1.14 eV.
However, the barrier vanishes if the final state is by $\sim$2 eV more
favorable than the initial one, or increases above 2 eV for diffusion in the
opposite direction. For most of the configurations investigated in the present
study the activation energies for diffusion of atomic H are in the range from
0.7 to 1.5 eV. 
Applying the linear least square fitting to the DFT values calculated for E$_a$ (the activation energy for H atom diffusion between 
adjacent C sites) and $\Delta E$ (the energy difference between configurations 
after and prior to H diffusion),  we arrive at the folowing expressions for clusters with $n$ H atoms:
\begin{equation} \label{eq-diff}
% E_a (n) = [1.1 -(-0.1)^n]\phantom{00} {\rm eV} + 0.5 \Delta E, 
 E_a (n) =\left\{\begin{array}{ll}
 1.0\>\mathrm{eV} + 0.5 \Delta E, &n \phantom{0}\mathrm{even}\\
 1.2\>\mathrm{eV} + 0.5 \Delta E, &n \phantom{0}\mathrm{odd}.
\end{array}
\right.
\end{equation}
We did not find clear arguments for observed energy shift of 0.2 eV in the 
E$_a$ between configurations with even (dimers, tetramers and hexamers) and 
odd (trimers and pentamers) number of H atoms.
Yet, Eq. \ref{eq-diff} and the BEP plot in Fig.\ \ref{bep} demonstrate 
that for H structures on graphene a fairly accurate estimate of their 
stability against H diffusion can be obtained from total energies of initial 
and final states, without explicit calculation of the corresponding activation 
energy E$_{\rm a}$.\\
%
% Opposite configurations
%
\section{Conclusions}\label{conc}
We investigated an extensive set of adsorption configurations of H atom
clusters on graphene, the H atoms being either all on one side of the graphene (\textit{cis}-clusters) 
or on both sides in an alternating manner (\textit{trans}-clusters). The binding energy per H atom in general increases with
the size of the clusters. The value of 0.77, calculated for monomers, 
increases to 1.6 eV in cis-hexamers and to 2.08 eV in trans-hexamers. H-H
interactions appear to favor cis-cluster shapes having H atoms in O- and P-positions with respect to each other
and to favor trans-clusters having H atoms in O-positions with respect to each other.
Very stable trans-clusters with 13-22 H atoms were identified by optimizing the number of H atoms in 
ortho-trans-positions and thereby the number of closed, H-covered carbon hexagons. For such clusters, H binding energies up to 2.23 eV were found.
For the cis-clusters, associative H$_2$ desorption
was investigated. It was found that desorption occurred from
P-dimers embedded in the clusters or from pairs of H atoms that first
rearranged via diffusion into embedded P-dimers, with barriers ranging
from $\sim$1.4 to $\sim$1.7 eV.
Upon formation of clusters with an odd number of
adsorbates a non-zero spin-density is induced. However, for the most stable
cis-configurations the spin-polarization is rather weak. The most stable 
cis-configurations with an even number of H atoms are all non-magnetic.
Yet, some of the very stable trans-clusters are magnetic, which opens routes 
for design of magnetic materials based on graphene functionalized with 
hydrogen.\\
This work has been supported by the Serbian Ministry of Science and
Technological Development under Grant No.\ 141039A and by the Danish Research Councils.
The calculations were performed at the Danish Center for Scientific Computing.

\end{document}